\documentclass[a4paper,10pt]{article}
\usepackage[utf8]{inputenc}
\usepackage{mathrsfs}
\usepackage{graphicx}
\usepackage{amsthm,amsfonts,amsmath} 
\newtheorem{prop}{Proposition}[section]

\newcommand{\p}{\partial}
\newcommand{\be}{\begin{equation}}
\newcommand{\ee}{\end{equation}}
\newcommand{\la}{\label}
\newcommand{\ba}{\begin{align}}
\newcommand{\ea}{\end{align}}
\begin{document}
\begin{center}
\section*{The Einstein  metrics with smooth scri}
\end{center}

-\centerline{J. Tafel}

\noindent
\centerline{Institute of Theoretical Physics, University of Warsaw,}
\centerline{Pasteura 5, 02-093 Warsaw, Poland, email: tafel@fuw.edu.pl}

\bigskip

\begin{abstract}
 We consider  solutions of the Einstein equations with cosmological constant $\Lambda\neq 0$  admitting conformal compactification with  smooth  scri $\mathscr{I^+}$. Metrics are written in the Bondi-Sachs coordinates  and expanded into inverse powers of the affine distance $r$. Unlike in the case $\Lambda=0$ all free data are located on the scri. There  are linear differential  constraints  on the Bondi mass and angular momentum aspects. All other components of metrics are defined in a recursive way. 
\end{abstract}

\bigskip
\section{Introduction}
This paper is a continuation of \cite{t} where the case $\Lambda=0$ is considered. We assume that metric $\tilde g$ satisfying vacuum equations with $\Lambda\neq 0$ is conformally equivalent to metric $g$ which is  smooth in a neighbourhood  of the scri $\mathscr{I^+}$. 
We consider its Bondi-Sachs form with the affine distance $r$. Because of this choice our approach slightly differs from the standard one using the luminosity distance (see \cite{bt,c} and references).  If metric is expanded into inverse powers of $r$ almost all Einstein equations  can be solved in an algebraic way with respect to metric coefficients. Only the  mass aspect $M$ and the angular momentum aspect $L_A$ undergo some linear differential constraints on $\mathscr{I^+}$. All free data are defined on the scri. The constraints on $M$ and $L_A$ reduce to a single Laplace-Beltrami equation in special coordinates. If metric tends asymptotically to the (anti) de Sitter solution this equation implies preservation of the generalized Bondi energy and  the oscilatory (for $\Lambda<0$) or 
exponential (for $\Lambda>0$) behaviour of the  linear momentum.

\section{Recursive solvability of the Einstein equations}
Let ($M,g$) be a partial conformal compactification of spacetime ($\tilde M,\tilde g$) with
a boundary $\mathscr{I^+}=R\times \Sigma$ foliated by 2-dimensional spacelike surfaces. 
In the standard way \cite{t} in this neighbourhood we introduce a system of the Bondi-Sachs coordinates $u$, $\Omega=1/r$, $x^A$ such that 
\be\la{1}
g=du(g_{00}du-2d\Omega+2 g_{0A}dx^A)+g_{AB}dx^Adx^B\ ,
\ee
where
\be\la{2}
\hat g_{0A} =0
\ee
(the hat  denotes value on $\mathscr{I^+}$). We assume that components $g_{\mu\nu}$ admit the Taylor expansion in $\Omega$ of  any order. 
 The physical metric is given by
 \be\la{3}
\tilde g=\Omega^{-2} g=du(\tilde g_{00}du+2dr+2\tilde g_{0A}dx^A)+\tilde g_{AB}dx^Adx^B\ ,
\ee
where $\tilde g_{00},\tilde g_{AB}$ are of the order $r^2$ and $\tilde g_{0A}=O(r)$. 

 We impose on (\ref{3})  the Einstein equations with  cosmological constant $\Lambda\neq 0$
 \be\la{6a}
 \tilde R_{\mu\nu}+\Lambda\tilde g_{\mu\nu}=0\ .
 \ee
 In terms of the compactified metric they are
 \be\la{7}
 R_{\mu\nu}-2Y_{\mu\nu}-g_{\mu\nu}Y^{\ \alpha}_{\alpha}=0 \ ,
 \ee
 where
 \be\la{8}
Y_{\mu\nu}=-\frac{1}{\Omega}\Omega_{|\mu\nu}+\frac {1}{2\Omega^{2}}(\Omega_{|\alpha}\Omega^{|\alpha}-\frac{\Lambda}{3})g_{\mu\nu}
 \ee
 and ${}_{|\mu}$ denotes the covariant derivative defined by $g$.
 
 Assume that $\tilde R_{\mu\nu}+\Lambda\tilde g_{\mu\nu}$ is finite on  $\mathscr{I^+}$. Then  $Y_{\mu\nu}$ must be also finite. In order to avoid a second order pole at  $\Omega=0$ one has to assume that
 \be\la{10}
g_{00}=-\frac{\Lambda}{3}+a\Omega+b\Omega^2-2M\Omega^3+ O(\Omega^4)\ .
\ee
 Thus, metric  induced by $g$ on the scri reads
 \be\la{8b}
 \hat g=-\frac{\Lambda}{3}du^2+\hat g_{AB}dx^Adx^B
 \ee
 and $\mathscr{I^+}$ is spacelike if $\Lambda>0$ and timelike if $\Lambda<0$. For the (A)dS metric there is $\hat g_{AB}=-s_{AB}$. Still we are free to change coordinates $u,x^A$ provided that $\hat g$ preserves form (\ref{8b}) modulo a conformal factor.
  It means that we can impose one condition on  $\hat g_{AB}$, e. g. 
we can try   to obtain 
\be\la{11f}
\det{\hat g_{AB}}=f\ ,\ \ f_{,0}=0
\ee
with a prescribed function $f$. 
For instance, if  $\mathscr{I^+}=R\times  S_2$, where $S_2$ is the 2-dimensional sphere, a natural condition would be
\be\la{11g}
\det{\hat g_{AB}}=\det{s_{AB}}\ .
\ee
Equation (\ref{11g}) is a reminiscent of the Bondi luminosity gauge
\be\la{11h}
\det{g_{AB}}=\det{s_{AB}}\ .
\ee

Regularity of $Y_{1A}$ yields
\be\la{10a}
g_{0A}=q_A\Omega^2+2L_A\Omega^3+O(\Omega^4)\ .
\ee
A lack of singularities in  $Y_{AB}$ is equivalent to the relation
\be\la{11}
\frac{\Lambda}{3}n_{AB}=\hat g_{AB,0}-a\hat g_{AB}
\ee
between expansion coefficients of  $g_{AB}$
\be\la{12}
g_{AB}=\hat g_{AB}+n_{AB}\Omega +p_{AB}\Omega^2+O(\Omega^3)\ .
\ee
All other components of  $Y_{\mu\nu}$ are nonsingular at $\Omega=0$ under conditions (\ref{10})-(\ref{11}).

Unlike in the case $\Lambda=0$, equation (\ref{11})  defines $n_{AB}$ in terms of $\hat g_{AB}$ and $a$. This relation can be further simplified by means of  a shift of $r$ leading to 
\be\la{12a}
a=0\ .
\ee
We will not  assume (\ref{12a}) at this stage  since another condition 
\be\la{12b}
n=0\ ,
\ee
where $n=\hat g^{AB}n_{AB}$, may be more convenient. Note that (\ref{12b}) is equivalent to 
\be\la{12c}
a=\frac 12(\ln{\det\hat g_{AB}})_{,0}
\ee
thanks to (\ref{11}).
Note that under (\ref{11f})  conditions (\ref{12a}) and (\ref{12b}) are equivalent. 
We will test  usefulness of the above  conditions  after obtaining field equations.

Folowing \cite{t} we can reduce a number of the Einstein equations (\ref{6a}) by means of the Bianchi identity. To this end it is sufficent to replace  $\tilde R_{\mu\nu}$ by $\tilde R_{\mu\nu}+\Lambda\tilde g_{\mu\nu}$ and $\tilde G_{\mu \nu}$ by $\tilde G_{\mu \nu}-\frac 12\Lambda\tilde g_{\mu\nu}$ in equation (24) in \cite{t}. In this way one obtains a minimal system of equations given by 
\be\la{40a}
  \tilde R_{11}^{(k)}=\tilde R_{1A}^{(k)}=0\ ,\ \ (\tilde R_{AB}+\Lambda \tilde g_{AB})^{(k)}=0
  \ee
  and
\be\la{40b}
   (\tilde R_{00}+\Lambda \tilde g_{00})^{(2)}=(\tilde R_{0A}+\Lambda \tilde g_{0A})^{(2)}=0\ ,
   \ee
where the subscript $(k)$ denotes coefficient of order $k$ in an expansion with respect to $\Omega$.

 Equation $\tilde R_{11}=0$ and  $\tilde R_{1A}=0$ are similar in character to those for $\Lambda=0$. They yield
 \be\la{a2a}
 \hat g^{AB}g^{(k+2)}_{AB}=\langle g^{(l)}_{AB},l\leq k+1\rangle\ ,\ k\geq 0\ ,
\ee
 \be\la{a4a}
g^{(k+2)}_{0A}=\langle g^{(l)}_{\mu\nu},l\leq k+1\rangle\ ,\ k\geq 0\ ,\  k\neq 1\ ,
\ee
where brackets on the r. h. s. denote expressions depending on functions within the bracket.
For $k=0$ equation (\ref{a4a})  reads
\be\la{a5}
q_A=-\frac 12n^B_{\ A|B}+\frac 12n_{,A}\ .
\ee
Equation  $\tilde R_{1A}^{(1)}=0$ does not define  $g^{(3)}_{0A}$. It is equivalent to 
\be\la{a6}
(p^B_{\ A}+\frac 14 nn^A_{\ B}-\frac 12 n^{BC}n_{AC})_{|B}+\frac 18(n_{BC}n^{BC}- n^2)_{,A}=0\ ,
\ee
where operations on indices and covariant derivatives are defined by  $\hat g_{AB}$. 
For $\Lambda=0$ we used in \cite{t} the topology of  $\mathscr{I^+}=R\times  S_2$ to show that (\ref{a6}) yields 
\be\la{a7}
p_{AB}=\frac 18(n_{AB}n^{AB}-n^2)\hat g_{AB}+\frac 14nn_{AB}\ .
\ee
 For $\Lambda\neq 0$ 
equation (\ref{a7}) follows algebraically from $(\tilde R_{AB}+\Lambda \tilde g_{AB})^{(0)}=0$. The latter equation also implies 
\be\la{22}
b=-\frac 12\hat R+\frac 12n_{,0}+\frac 14an+\frac{\Lambda}{8}(n_{AB}n^{AB}-\frac 13n^2)\ ,
\ee
where $\hat R$ is the Ricci scalar of $\hat g_{AB}$. From the point of view  of equations (\ref{a7}) and (\ref{22}) the most convenient gauge condition is (\ref{12b}). In this gauge $n_{AB}$ is proportional to the traceless part of the exterior curvature  of the section  $u=const$ of $\mathscr{I^+}$ and $p_{AB}$ is proportional to  metric of this section.

Equation $(\tilde R_{AB}+\Lambda \tilde g_{AB})^{(k)}=0$ with $k=1$ does not contain any new information. For $k\geq 2$ it can be splitted into its trace (with respect to $\hat g_{AB}$) and a traceless part. The trace part is equivalent to
 the following composition of equations 
\be\la{a10}
\hat g^{AB}(\tilde R_{AB}+\Lambda \tilde g_{AB})^{(k)}-\frac{\Lambda (k-3)}{3(k+1)}\tilde R^{(k)}_{11}=0\ ,\ k\geq 2\ .
\ee
It allows to obtain $g_{00}^{(k+2)}$ in terms of  lower order coefficients 
\be
g^{(k+2)}_{00}=\langle g^{(l)}_{\mu\nu},l\leq k+1\rangle\ ,\ k\geq 2\ .\la{a11}
\ee
The traceless part of $(\tilde R_{AB}+\Lambda \tilde g_{AB})^{(k)}=0$ yields the traceless part of $g_{AB}^{(k)}$ (denoted by a ring)
\be\la{a12}
\mathring g^{(k+2)}_{AB}=\langle g^{(l)}_{\mu\nu},l\leq k+1\rangle\ ,\ k\geq 2\ .
\ee 
Note that in the case $\Lambda=0$ instead of (\ref{a12}) one obtains an expression for $\mathring g^{(k+1)}_{AB,0}$ in terms of lower order coefficients \cite{t}. Thus, (\ref{a12}) is the second equation, after (\ref{11}), which makes a qualitative difference between $\Lambda=0$ and $\Lambda\neq 0$. For $\Lambda\neq 0$ there is no need of initial data for $g_{AB}$.   Moreover, tensor 
\be\la{53a}
N_{AB}=\mathring g_{AB}^{(3)}
\ee
(instead of $n_{AB}$) is arbitrary if $\Lambda\neq 0$. 

 The last equations to consider are (\ref{40b}). They are equivalent to equations
 \be\la{a22}
 (\tilde R_{00}+\Lambda \tilde g_{00})^{(2)}-\frac{\Lambda}{6}\hat g^{AB}(\tilde R_{AB}+\Lambda \tilde g_{AB})^{(2)}-\frac{\Lambda^2}{18}\tilde R _{11}^{(2)}-\frac{\Lambda}{9}a\tilde R_{11}^{(1)}=0\ ,
 \ee
\be\la{b22}
(\tilde R_{0A}+\Lambda \tilde g_{0A})^{(2)}-\frac{\Lambda}{3}\tilde R_{1A}^{(2)}+\frac {\Lambda}{6}\tilde R_{11,A}^{(1)}=0
\ee
having the following structure
 \be\la{22a}
M_{,0}+\frac{3}{4}(\ln{|\hat g|})_{,0}M+\frac{\Lambda}{2}L^A_{\ |A}=\frac{\Lambda^2}{24}n^{AB} N_{AB}+\langle a,\hat g_{AB}\rangle\ ,
\ee
 \be\la{22b}
L_{A,0}+\frac 12(\ln{|\hat g|})_{,0}L_A+\frac 13M_{,A}=\frac{\Lambda}{6} N_{AB}^{\ \ |B}+\langle a,\hat g_{AB}\rangle\ .
\ee
Given $a,\ \hat g_{AB}$ and $N_{AB}$ equations (\ref{22a}) and (\ref{22b}) constitute a conjugate system of linear equations for $M$ and $L_A$.
Condition (\ref{11f}) allows to simplify them to
\be\la{a22a}
M_{,0}+\frac{\Lambda}{2}L^A_{\ |A}=f\ ,
\ee
\be\la{a22b}
L_{A,0}+\frac 13M_{,A}=h_A\ ,
\ee
where $f$ and $h_A$ are known functions. Let us introduce function $ M'$ such that 
\be\la{53}
M=M'_{,0}\ .
\ee
Then (\ref{a22b}) yields
\be\la{54}
L_A=-\frac 13 M'_{,A}+\int{h_Adu}
\ee
and (\ref{a22a}) becomes 
\be\la{55}
-\frac{6}{\Lambda}M'_{,00}+\hat \Delta M'=-\frac{6}{\Lambda}f+3\nabla^A\int{h_Adu}\ ,
\ee
where $\hat \Delta$ is the covariant Laplace operator with respect to $\hat g_{AB}$. Thus, equations (\ref{22a}) and (\ref{22b}) reduce to a hyperbolic (if $\Lambda<0$) or elliptic ($\Lambda>0$) equation for $M'$.
Note that the operator acting on $M'$ is not  the Laplace-Beltrami operator of the induced  metric (\ref{8b})  on $\mathscr {I}^+$  because of the ``wrong`` coefficient $\frac{6}{\Lambda}$.

Let us  summarize  consequences of the  Einstein equations  expanded into powers of $1/r$.
\begin{prop}
The Einstein metric with $\Lambda\neq 0$ and smooth scri $\mathscr{I^+}$ is given by 
\be\la{3a}
\tilde g=du(\tilde g_{00}du+2dr+2\tilde g_{0A}dx^A)+\tilde g_{AB}dx^Adx^B\ ,
\ee 
\be\la{a10d}
\tilde g_{00}=-\frac{\Lambda}{3}r^2+ar+b-\frac{2M}{r}+\Sigma_2^{\infty} g_{00}^{(k+2)}r^{-k}\ ,
\ee
\be\la{a10a}
\tilde g_{0A}=q_A+\frac{2L_A}{r} +\Sigma_2^{\infty} g_{0A}^{(k+2)}r^{-k}\ ,
\ee
\be\la{a12c}
\tilde g_{AB}=r^2\hat g_{AB}+rn_{AB} +p_{AB}+\frac 1r(N_{AB}+N\hat g_{AB})+\Sigma_2^{\infty} g_{AB}^{(k+2)}r^{-k}\ ,
\ee
where  coefficients are  defined  in the following way:
\begin{itemize}
 \item u-dependent metric $\hat g_{AB}$ and a traceless (with respect to $\hat g_{AB}$) tensor $N_{AB}$ can be arbitrarily chosen up to a gauge  condition, e.g. (\ref{11f}). 
 \item Function $a$ is arbitrary but it can be gauged away, e.g. by means of  (\ref{12c}).
 \item Components $n_{AB}$, $p_{AB}$, $N$, $q_A$ and $b$   are defined by $a$ and $\hat g_{AB}$  according to (\ref{11}), (\ref{a7}), (\ref{a2a}), (\ref{a5}) and (\ref{22}).
\item Coefficients $M$ and $L_A$ are defined by their initial values at $u=u_0$ via  equations (\ref{22a}) and (\ref{22b}). Under condition (\ref{11f}) these equations  reduce to  equation (\ref{55}) for $M'$ and relations (\ref{53}) and (\ref{54}).
\item All other components are defined in a recursive way by equations (\ref{a2a}), (\ref{a4a}), (\ref{a11}) and (\ref{a12}).
\end{itemize}
\end{prop}
The main difference between present situation and that for $\Lambda=0$ (see Proposition 2.1 in \cite{t}) is that now all free data are located at $\mathscr{I^+}$. It is interesting that for $\Lambda<0$ there is no need of the Cauchy data ($\mathscr{I^+}$ is timelike). 
Evolution of the Cauchy  data given on $\mathscr{I^+}$  for $\Lambda>0$ was first obtained by Friedrich \cite{f}.

\section{Solutions  with the (A)dS boundary metric} 
If metric $\tilde g$ tends asymptotically to the (anti) de Sitter solution   then
\be\la{56}
\hat g_{AB}=-s_{AB}\ .
\ee
 In this case it follows from  equations (\ref{10})-(\ref{11}) and (\ref{a2a})-(\ref{22}) that in the gauge $a=0$ there is 
 \be\la{56a}
 \tilde g_{00}=-\frac{\Lambda}{3}r^2+1-\frac{2M}{r}+O(\frac{1}{r^2})\ ,
 \ee
 \be\la{56b}
 \tilde g_{0A}=\frac{2L_A}{r}+O(\frac{1}{r^2})\ ,
 \ee
 \be\la{56c}
 \tilde g_{AB}=-r^2s_{AB}+\frac 1r N_{AB}+O(\frac{1}{r^2})\ .
 \ee
 The r. h. s. of equations (\ref{a22a}) and (\ref{a22b}) is given by 
\be\la{57}
h_A=\frac{\Lambda}{6} N^{\ \ |B}_{AB}\ ,\ f=0\ .
\ee
Let us introduce a  tensor $N'_{AB}$ such that 
\be\la{58}
N_{AB}=N'_{AB,0}\ ,\ \ s^{AB}N'_{AB}=0\ .
\ee
Equation  (\ref{54}) yields 
\be\la{59}
L_A=-\frac 13 M'_{,A}+\frac{\Lambda}{6}{N'}_{AB}^{\ \ |B}+\tilde L_A\ ,\ \ \tilde L_{A,0}=0\ .
\ee
Using decomposition
\be\la{59a}
\tilde  L_A=f_{,A}+\eta^C_{\ A}h_{,C}
\ee
we can incorporate $f$ into $M'$ and $h$ into $N'_{AB}$  except the dipole component of $h$ (see Lemma 3.1 in \cite{t}). Thus, without loss of generality we can assume that 
\be\la{59b}
\tilde  L_A=\eta^C_{\ A}h_{,C}\ ,\ \ h=\alpha^kY_{1k}\ ,
\ee
where $Y_{lk}$ are spherical harmonics and $\alpha^k=const$.
In spherical coordinates $\theta,\varphi$ related  to the axis defined by   $\alpha^k$ one obtains expression
\be\la{60a}
\tilde L_Adx^A=\alpha\sin^2{\theta}d\varphi\ ,
\ee
which is the angular momentum term known from the Kerr solution.
In view of (\ref{59}) and (\ref{59b})  equation (\ref{a22a}) transforms into
\be\la{60}
\frac{6}{\Lambda}M'_{,00}+\Delta M'=-\frac {\Lambda}{2} {N'}^{AB}_{\ \ \ |AB}\ ,
\ee
where $\Delta$ is the standard Laplace operator on the sphere.

 If we integrate equation (\ref{60}) over the sphere we obtain the known result
\be\la{81}
\oint_{S_2}{Md\sigma}=const.
\ee
The l. h. s. of (\ref{81}) is,   modulo $4\pi$,  rather unique  candidate for the total energy  in this case. One can interpret  (\ref{81})
as a lack of gravitational radiation.
If we integrate (\ref{60}) with $Y_{1k}$ we get equation
\be\la{85}
\frac{6}{\Lambda}M'_{k,00}-2 M'_k=0
\ee
for dipole moments
\be\la{84}
M'_k=\oint_{S_2}{M'Y_{1k}d\sigma}\ .
\ee
The Bondi  linear  momentum $P_k$ is proportional to the time derivatives of $M'_k$, so it satisfies equation
\be\la{86}
P_{k,00}=\frac{\Lambda}{3}P_k\ .
\ee
Hence, it  either oscilates when $u$ changes (for $\Lambda<$) or behaves in an exponential way (for $\Lambda>0$). In the second case its square must exceed the total energy either for increasing or decreasing $u$, so a reasonable physical assumption would be $M_k=0$. 
Higher moments of $M'$ satisfy nonhomogeneous equations coming from (\ref{60}). Their solutions are defined in quadratures up to  oscillatory or exponential functions depending on sign of $\Lambda$.

An alternative approach to equation  (\ref{60}) is to treat it as an equation for  tensor $N'_{AB}$. We can represent this tensor  
by two scalar functions $f$ and $h$ \cite{t}
\be\la{62}
N'_{AB}=f_{|AB}-\frac 12 \Delta fs_{AB}+\eta^C_{\ (A}h_{|B)C}\ .
\ee
Substituting (\ref{62}) into (\ref{60}) yields
\be\la{63}
\frac{6}{\Lambda}M'_{,00}+\Delta M'=-\frac {\Lambda}{4} (\Delta+2)\Delta f\ .
\ee
Function $f$ exists if conditions (\ref{81}) and (\ref{84}) are satisfied.
Then we can write 
\be\la{66a}
f=-\frac{4}{\Lambda}\Delta^{-1}(\Delta+2)^{-1}(\frac{6}{\Lambda}M'_{,00}+\Delta M')\ .
\ee 
Formula (\ref{66a}) can be realised by decomposition of $M'$ into $Y_{lk}$. In this approach functions $M'$ and $h$ are arbitrary modulo conditions (\ref{81}) and (\ref{84}).
We summarize above results in the following proposition.
\begin{prop}
The Einstein metric which has a smooth scri $\mathscr{I^+}$ and tends asymptotically to the (A)dS metric is given by (\ref{3a}) and (\ref{56a})-(\ref{56c}). 
Components $M$, $N_{AB}$ and $L_A$  are defined by solution $(M',N'_{AB})$ of equation (\ref{60}) via relations (\ref{53}), (\ref{58}), (\ref{59}) and (\ref{60a}).  Higher order  components
 are defined in a recursive way by equations (\ref{a2a}), (\ref{a4a}), (\ref{a11}) and (\ref{a12}). The Bondi energy is constant and the Bondi linear momentum satisfies (\ref{86}).
\end{prop}

\section{Stationary merics} 

If metric considered in section 2  admits a timelike Killing vector $K$ then we can transform it to  the form (\ref{3a}) with $u$-independent coefficients  but we cannot assume (\ref{2}) (see Section 3 in \cite{t})
The KIlling vector is given by  $K=\p_0$. 
Let us assume (\ref{12}) and expansions
 \be\la{62a}
g_{00}=\hat g_{00}+a\Omega+b\Omega^2-2M\Omega^3+ O(\Omega^4)\ ,
\ee
\be\la{62b}
g_{0A}=\hat g_{0A}+v_{A}\Omega+q_A\Omega^2+2L_A\Omega^3+O(\Omega^4)
\ee
with coefficients depending  on coordinates $x^A$.
Tensor  (\ref{8})  is finite on the boundary iff
\be\la{63}
\hat g_{00}=-\frac{\Lambda}{3}+\hat g_{0A}\hat g_0^{\ A}\ ,
\ee
\be\la{63a}
v_A=n_A^{\ B}\hat g_{0B}
\ee
and
\be\la{64}
\frac{\Lambda}{3}n_{AB}=-2\hat g_{0(A|B)}+(n_{CD}\hat g_0^{\ C}\hat g_0^{\ D}-a)\hat g_{AB}\ ,
\ee
where indices $A,B$ are raised by means of $\hat g^{AB}$. Since $\hat g_{0A}\hat g_0^{\ A}\leq 0$ and we exclude $\Lambda=0$ it follows from (\ref{63})-(\ref{64}) and timelike property of $K=\p_0$ that 
\be\la{72}
\Lambda<0\ ,\ \ |\hat g_{0A}\hat g_0^{\ A}|\leq \frac{|\Lambda|}{3}\ .
\ee
Thus, metrics are of the AdS type.

 It is convenient to shift  coordinate $r$  in order to obtain 
\be\la{64c}
n^A_{\ A}=0\ .
\ee
Equation (\ref{64}) is then  equivalent to the following relations defining $n_{AB}$ and $a$ 
\be\la{64b}
\frac{\Lambda}{3}n_{AB}=-2\hat g_{0(A|B)}+\hat g_{0\ |C}^{\ C}\hat g_{AB}\ ,
\ee
\be\la{64a}
a=n_{CD}\hat g_0^{\ C}\hat g_0^{\ D}-\hat g_{0\ |A}^{\ A}\ .
\ee
Thus, again $n_{AB}$ is proportional to the traceless part of the exterior curvature  of the section  $u=const$ of $\mathscr{I^+}$.

Transformation of coordinates $x^A$ allows us to obtain
\be\la{64d}
\hat g_{AB}=-\gamma^2s_{AB}\ ,
\ee
where $\gamma$ is a positive function of $x^A$. Still there is a freedom of supertranslation  
\be
u\rightarrow u+f(x^A)
\ee
which can be used to impose a condition on $\hat g_{0A}$, $\gamma$ or higher order coefficients in metric $g$. 
All the Einstein equations except (\ref{40b})
can be solved algebraically   as in the general case in Section 2.
Equations (\ref{40b})
are now a  complicated  system of constraints on $M$,  $L_A$, $\gamma$, $\hat g_{0A}$ and  $N_{AB}$, which is linear in  $M$,  $L_A$ and $N_{AB}$. 

The space of stationary solutions is much bigger than in the case  $\Lambda=0$ \cite{t}. In general, solutions do not tend asymptotically to the AdS metric. It is so if
the boundary metric 
\be\la{65}
\hat g=-\frac{\Lambda}{3}du^2-\gamma^2 s_{AB}(dx^A+\hat g_0^{\ A}du)(dx^B+\hat g_0^{\ B}du)
\ee
is conformally equivalent to 
\be\la{66}
\hat g'=-\frac{\Lambda}{3}du'^2-s_{AB}dx'^Adx'^B\ .
\ee
Metric (\ref{66}) is conformally flat (note that it coincides with the  conformal compactification of the 3-dimensional Minkowski metric). Thus, 
metric  (\ref{65}) must be also conformally flat. Since $K=\p_u$ is the Killing field of (\ref{65}) it is  a conformal Killing vector  of the  flat metric. Such vectors compose   the  10-dimensional algebra $so(2,3)$. 
Given one of them  in terms of the Cartesian coordinates of flat space one can find (at least in principle) coordinates $u,\ x^A$ such that $K=\p_u$. Writing flat metric in these coordinates shows what   metrics (\ref{65}) are available 
if the physical metric  $\tilde g$ is asymptotically AdS (in transformed coordinates). 

\section{Summary}
In this paper we considered the Einstein metrics  admitting a conformal compactification with smooth scri. For nonvanishing cosmological constant the scri is either spacelike ($\Lambda>0$) or timelike ($\Lambda<0$). Metrics can be transformed to  the Bondi-Sachs form with components which can be  expanded into powers of a radial distance. The Einstein equations imply that the expansion  coefficients  can be expressed by coefficients of lower order (see Proposition 2.1). Beside these recursive formulas there are 3 differential conditions constraining  the analog of the Bondi mass and angular momentum aspect. For a special foliation of the scri they  can be reduced to one second order equation for one function.  All free data are located on the scri.

Asymptotic form of metric  simplifies considerably if it tends  to the (anty) de Sitter solution (see Proposition 3.1). In this case the total energy calculated according to the Bondi prescription is constant and the total momentum oscilates or behaves exponentially.

Finally we considered solutions with the timelike Killing field. In this case, in contrary to $\Lambda=0$, there is no big difference in solving the Einstein equations with respect to the general case. 
However, existence of stationary solutions with nonvanishing asymptotic twist (then $\hat g_{0A}\neq 0$) can be important for  a notion of  the total energy in general case with $\Lambda< 0$. 
We expect that a crucial ingredient in any definition of the energy should be the  mass aspect $M$ related to coordinates such that  equations (\ref{8b}) and (\ref{11h}) are satisfied. If
the timelike Killing field $K$  is present then transformation between coordinates used in (\ref{8b}) and those in (\ref{65})  must depend on time $u$. It seems unavoidable that the total energy also depend on time, what seems unacceptable in a stationary spacetime.

\null

\noindent
\textbf{Acknowledgments}

\null

\noindent
This work was partially   supported by Project OPUS 2017/27/B/ST2/02806
of Polish National Science Centre (NCN).

\end{document}